\documentclass[11pt]{article}
\def\be {\begin{equation}}
\def\ee {\end{equation}}
\def\bea {\begin{eqnarray}}
\def\eea {\end{eqnarray}}
\begin{document}
\title{{\bf{\Large Hawking radiation and black hole spectroscopy in Ho\v{r}ava-Lifshitz gravity}}}
\author{
 {\bf {\normalsize Bibhas Ranjan Majhi}$
$\thanks{E-mail: bibhas@bose.res.in}}\\
 {\normalsize S.~N.~Bose National Centre for Basic Sciences,}
\\{\normalsize JD Block, Sector III, Salt Lake, Kolkata-700098, India}
\\[0.3cm]
}

\maketitle
\begin{abstract}
     Hawking radiation from the black hole in Ho\v{r}ava-Lifshitz gravity is discussed by a reformulation of the tunneling method given in \cite{Banerjee:2008sn}. Using a density matrix technique the radiation spectrum is derived which is identical to that of a perfect black body. The temperature obtained here is proportional to the surface gravity of the black hole as occurs in usual Einstein gravity. The entropy is also derived by using the first law of black hole thermodynamics. Finally, the spectrum of entropy/area is obtained. The latter result is also discussed from the viewpoint of quasi-normal modes. Both methods lead to an equispaced entropy spectrum, although the value of the spacing is not the same. On the other hand, since the entropy is not proportional to the horizon area of the black hole, the area spectrum is not equidistant, a finding which also holds for the {\it Einstein-Gauss-Bonnet theory}.    
\end{abstract}
\section{Introduction}
    Inspired by condensed matter models of dynamical critical system, Ho\v{r}ava proposed a new four dimensional theory of gravity \cite{Horava:2009uw}, popularly known as Ho\v{r}ava-Lifshitz gravity. 
Since then a lot of attention has been given in several directions \cite{Horava:2008ih,Myung:2009ur} of this Ho\v{r}ava-Lifshitz gravity theory.

     Recently, it has also been shown that a static spherically symmetric black hole solution exists in this theory for the Lifshitz point $z=3$ \cite{Lu:2009em,Cai:2009pe} and for $z=4$ \cite{Cai:2009ar} {\footnote{For other black hole solutions of Ho\v{r}ava-Lifshitz gravity with different conditions see \cite{Colgain:2009fe,Harada:2009du}}}. A study of the thermodynamic properties of this black hole has also been done \cite{Cai:2009pe,Cai:2009qs,Cai:2009ph,Cai:2009ar,Myung:2009dc}. Surprisingly, however, there does not exist any detailed study of the {\it {Hawking effect}} \cite{Hawking:1974sw}, except for some sporadic attempts \cite{Peng:2009uh,Chen:2009bj}. The motivation of this paper is to fill such a gap. We feel this study to be important since the Hawking radiation is crucial to give the black holes one of its thermodynamic properties making it consistent with the rest of physics.

   The Hawking effect can be discussed by Hawking's original approach \cite{Hawking:1974sw} or anomaly method \cite{Robinson:2005pd,Banerjee:2007qs}. Here we will discuss the Hawking effect by a physically intuitive picture -  a reformulation \cite{Banerjee:2008sn,bm2} of the standard {\it tunneling formalism} {\footnote{For more references on tunneling mechanism see \cite{Majhi4}}} \cite{Paddy,Wilczek,Banerjee:2008ry}. The advantage of this approach \cite{Banerjee:2008sn,bm2} is that, in contrast to \cite{Paddy,Wilczek,Majhi2,Majhi3,Singleton}, the spectrum is directly obtained instead of just the temperature.

       In this paper, we will study the propagation of scalar fields on the background of Ho\v{r}ava-Lifshitz black hole spacetime for $z=3$. 
Following the reformulation of the tunneling mechanism \cite{Banerjee:2008sn,bm2,Banerjee:2009pf} the explicit expressions of the left and right moving modes in the semi-classical limit (i.e. $\hbar\rightarrow 0$) as seen by an asymptotic observer will be given. 
Exploiting a density matrix technique the radiation spectrum will be derived. We find that the distribution function exactly matches with the black body radiation with a temperature proportional to the surface gravity of the black hole. This is a new result in the context of black holes in Ho\v{r}ava-Lifshitz gravity.

    Also, we will calculate the thermodynamic entities of the black hole. Using the first law of thermodynamics the expression for the entropy will be derived. In this case the entropy is not proportional to the area of the event horizon as happens in Einstein gravity; rather it has an extra additive logarithmic term involving the area.

   Another purpose of this paper is to study the nature of entropy/area spectrum of the Ho\v{r}ava-Lifshitz black hole. Here we will use two methods: tunneling method \cite{Banerjee:2009pf} and quasi-normal mode (QNM) method \cite{Hod:1998vk,Kunstatter,Maggiore:2007nq}. 
In both approaches the entropy (and not area) spectrum is seen to be equispaced.
We also observe that, although there is a discrepancy in the value of the entropy spacing in the tunneling and quasi normal mode approaches, the order of magnitude is same. The probable reason for such discrepancy is also discussed here.

    The organization of the paper is as follows. In section-2 a short discussion on the black hole solution of Ho\v{r}ava-Lifshitz gravity 
 and the expressions of the modes of the scalar particle as seen by an asymptotic observer are presented. The radiation spectrum is calculated in the next section. Explicit forms of the temperature and the entropy of the black hole are derived in section-4. Section-5 is devoted for the discussions on the entropy/area spectrum. Finally, we give the concluding remarks.        
\section{Tunneling in black hole in Ho\v{r}ava-Lifshitz gravity}

   The spherically symmetric static black hole solution at the Lifshitz point $z=3$ in Ho\v{r}ava-Lifshitz theory was obtained as \cite{Cai:2009pe},
\begin{eqnarray}
ds^2_H=-N^2(r)dt^2 + \frac{dr^2}{f(r)} + r^2 d\Omega^2_a
\label{solution1}
\end{eqnarray}
where $d\Omega^2_a$ is the line element for a two dimensional Einstein space with constant scalar curvature $2a$. Without loss of generality, one can take $a=0, \pm 1$ respectively.
The metric coefficients are given by \cite{Cai:2009pe},
\begin{eqnarray}
f(r) = a + x^2 -\alpha x^{(2\lambda\pm\sqrt{6\lambda - 2})/(\lambda - 1)},
\label{metric1}
\\
N(r) = x^{-(1+3\lambda\pm 2\sqrt{6\lambda-2})/(\lambda-1)} \sqrt{f(r)}
\label{metric2}
\end{eqnarray}
where $\alpha$ is the integration constant and $\lambda$ is the coupling constant, susceptible to quantum corrections \cite{Horava:2009uw}. Here $x=r\sqrt{-\Lambda}$ with $\Lambda$ is the cosmological constant. Now, for the above metric coefficients to be real, $\lambda$ must be greater than $\frac{1}{3}$. Hence in this case, as explained in \cite{Lu:2009em}, the cosmological constant $\Lambda$ is negative.

     Henceforth we will consider the $\lambda=1$ case, which is of particular interest. The metric coefficients can then be obtained from (\ref{metric1}) and (\ref{metric2}) after taking the limit $\lambda\rightarrow 1$. Here it must be noted that for the positive sign of (\ref{metric1},\ref{metric2}), this limit does not exist, while for the other sign, it exists. Therefore, for the negative sign, it leads to the following solutions:
\begin{eqnarray}
N^2(r)=f(r)=a+x^2-\alpha\sqrt{x}; \,\,\,\ x=r\sqrt{-\Lambda}~. 
\label{solution2}
\end{eqnarray}
This solution is asymptotically $AdS_4$ and has a singularity at $x=0$ if $\alpha\neq 0$. This singularity could be covered by the black hole horizon at $x_+$, the largest root of the equation $f(x_+)=0$. For $a=1$, equation (\ref{solution2}) reduces to that obtained in \cite{Lu:2009em}. The ADM mass of this black hole is given by \cite{Cai:2009pe},
%
%
\begin{eqnarray}
M =  \frac{p^2\mu^2\sqrt{-\Lambda}\Omega_a}{16}\alpha^2~.
\label{mass}
\end{eqnarray}
Since $\Lambda$ is negative \cite{Lu:2009em} and $\alpha$ is a real constant, the black hole mass $M$ is always positive and real.

    Now to find the modes of the particle as seen by the observer situated outside the event horizon, 
consider the massless Klein-Gordon (KG) equation $g^{\mu\nu}\nabla_\mu\nabla_\nu\phi=0$ under the background (\ref{solution1}). 
Proceeding in a similar way as presented in \cite{Banerjee:2008sn,bm2} we obtain that the modes which
are travelling in the ``in'' and ``out'' sectors of the black hole horizon are connected through the expressions
\begin{eqnarray}
&&\phi^{(R)}_{in} = e^{-\frac{\pi\omega}{\hbar \kappa}} \phi^{(R)}_{out}
\label{trans1}
\\
&&\phi^{(L)}_{in} =  \phi^{(L)}_{out},
\label{trans2}
\end{eqnarray}
where $\omega$ is the effective energy of the emitted particle as measured at infinity and $\kappa$ is the surface gravity defined by
\begin{eqnarray}
\kappa = \frac{1}{2} \frac{d f(r)}{dr} \Big|_{r=r_{+}}~.
\label{sgrav}
\end{eqnarray}
Here $L$ ($R$) refers to the ingoing (outgoing) mode. $r_+$ is the event horizon of the black hole (\ref{solution1}). In the set (\ref{trans1},\ref{trans2}), the left hand side modes of the equality represent the pair produced inside the black hole while those on the right hand side of equality represent the modes of that pair if one observes from outside the black hole. Since the physical quantities for the black hole are measured from the outside observer, we will always use this set of transformations in the subsequent analysis.

  Now, since the left moving mode travels towards the center of the black hole, its probability to go inside,
as measured by an external observer, is 
\begin{eqnarray}
P^{(L)}=|\phi^{(L)}_{in}|^2 = |\phi^{(L)}_{out}|^2=1
\label{Krus4}
\end{eqnarray}
where we have used (\ref{trans2}) to recast $\phi^{(L)}_{in}$ in terms of $\phi^{(L)}_{out}$
since measurements are done by an outside observer.
This shows that the left moving (ingoing) mode is trapped inside the black hole, as expected.
On the other hand the right moving mode, i.e. $\phi^{(R)}_{in}$, tunnels through the event horizon.
So proceeding in the similar way we obtain 
the tunneling probability as seen by an external observer as 
$P^{(R)}=|\phi^{(R)}_{in}|^2 = |e^{-\frac{\pi\omega}{\hbar \kappa}}\phi^{(R)}_{out}|^2
=e^{-\frac{2\pi\omega}{\hbar \kappa}}$,
i.e. there is a finite probability for the outgoing mode to cross the horizon.
    
\section{Radiation spectrum}
       In this section we will derive the emission spectrum from the black hole in the tunneling approach with the help of density matrix technique developed in \cite{bm2}. Now to find this spectrum, we first consider $n$ number of non-interacting virtual pairs that are created inside the black hole. Each of these pairs is represented by the modes defined in the left side of the equality of (\ref{trans1}, \ref{trans2}). Then the physical state of the system, observed from outside, is given by,
\begin{eqnarray}
|\Psi> = N \sum_n |n^{(L)}_{\textrm{in}}>\otimes|n^{(R)}_{\textrm{in}}>
 = N \sum_n e^{-\frac{\pi n\omega}{\hbar {\kappa}}}|n^{(L)}_{\textrm{out}}>\otimes|n^{(R)}_{\textrm{out}}>
\label{1.31}
\end{eqnarray}
where  use has been made of the transformations (\ref{trans1}) and (\ref{trans2}). Here $|n^{(L)}_{\textrm{out}}>$ corresponds to $n$ number of left going modes and so on while $N$ is a normalization constant which can be determined by using the normalization condition $<\Psi|\Psi>=1$.  This immediately yields, 
$N=\Big(\displaystyle\sum_n e^{-\frac{2\pi n\omega}{\hbar {\kappa}}}\Big)^{-\frac{1}{2}}$.
The sum will be calculated for both bosons and fermions. For bosons $n=0,1,2,3,....$ whereas for fermions $n=0,1$. With these values of $n$ we obtain the normalization constant as \cite{bm2},
\begin{eqnarray}
N_{(\textrm {boson})}=\Big(1-e^{-\frac{2\pi\omega}{\hbar {\kappa}}}\Big)^{\frac{1}{2}};\,\,\
\label{1.33}
N_{(\textrm {fermion})}=\Big(1+e^{-\frac{2\pi\omega}{\hbar \kappa}}\Big)^{-\frac{1}{2}}
\end{eqnarray}
From here on our analysis will be only for bosons since for fermions the analysis is identical. For bosons the density matrix operator of the system is given by,
\begin{eqnarray}
{\hat\rho}_{(\textrm{boson})}&=&|\Psi>_{(\textrm{boson})}<\Psi|_{(\textrm{boson})}
\nonumber
\\
&=&\Big(1-e^{-\frac{2\pi\omega}{\hbar \kappa}}\Big) \sum_{n,m} e^{-\frac{\pi n\omega}{\hbar \kappa}} e^{-\frac{\pi m\omega}{\hbar \kappa}} |n^{(L)}_{\textrm{out}}>\otimes|n^{(R)}_{\textrm{out}}>  <m^{(R)}_{\textrm{out}}|\otimes<m^{(L)}_{\textrm{out}}|
\label{1.37}
\end{eqnarray}
on exploiting (\ref{1.31}) with the normalization (\ref{1.33}).
Now tracing out the ingoing (left) modes we obtain the density matrix for the outgoing modes,
\begin{eqnarray}
{\hat{\rho}}^{(R)}_{(\textrm{boson})}= \Big(1-e^{-\frac{2\pi\omega}{\hbar \kappa}}\Big) \sum_{n} e^{-\frac{2\pi n\omega}{\hbar \kappa}}|n^{(R)}_{\textrm{out}}>  <n^{(R)}_{\textrm{out}}|
\label{1.38}
\end{eqnarray}
Therefore the average number of particles detected at asymptotic infinity is given by,
\begin{eqnarray}
<n>_{(\textrm{boson})}={\textrm{trace}}({\hat{n}} {\hat{\rho}}^{(R)}_{(\textrm{boson})})&=& \Big(1-e^{-\frac{2\pi\omega}{\hbar \kappa}}\Big) \sum_{n} n e^{-\frac{2\pi n\omega}{\hbar \kappa}}
\nonumber
\\
&=&\frac{1}{e^{\frac{2\pi\omega}{\hbar \kappa}}-1}
\label{1.39}
\end{eqnarray}
where the trace is taken over all $|n^{(R)}_{\textrm{out}}>$ eigenstates. This is the Bose distribution. Similar analysis for fermions leads to the Fermi distribution:
\begin{eqnarray}
<n>_{(\textrm{fermion})}=\frac{1}{e^{\frac{2\pi\omega}{\hbar \kappa}}+1}
\label{1.40} 
\end{eqnarray}
Note that both these distributions correspond to a black body spectrum with a temperature given by the Hawking expression, 
\begin{eqnarray}
T_H=\frac{\hbar \kappa}{2\pi}
\label{1.30}
\end{eqnarray}
Correspondingly, the Hawking flux can be obtained by integrating the above distribution functions over all $\omega$'s. For fermions it is given by,
\begin{eqnarray}
{\textrm{Flux}}=\frac{1}{\pi}\int_0^\infty \frac{\omega ~d\omega}{e^{\frac{2\pi\omega}{\hbar K}}+1} =\frac{\hbar^2 \kappa^2}{48\pi}
\label{1.41}
\end{eqnarray}
Similarly, the Hawking flux for bosons can be calculated, leading to the same answer.

\section{Black hole thermodynamics}
      In the previous section, the emission spectrum of the particle from the black hole has been derived. This is a perfectly black body spectrum with the temperature given by (\ref{1.30}). Now to find out the explicit form of the temperature, we will derive the surface gravity for this black. Use of the definition (\ref{sgrav}) and explicit form of metric coefficient (\ref{solution2}) yield the value of the surface gravity as,
\begin{eqnarray}
\kappa=\frac{1}{2}\Big(2x_+\sqrt{-\Lambda} - \frac{\alpha}{2}\sqrt{\frac{-\Lambda}{x_+}}\Big)
= \frac{3x_+^2 - a}{4x_+}\sqrt{-\Lambda}
\label{kappa}
\end{eqnarray} 
where, in the last step, we have substituted the value of $\alpha$ from the equation $f(x_+)=0$. Therefore substituting this in (\ref{1.30}) the Hawking temperature of the black hole is given by
\begin{eqnarray}
T_H=\frac{\hbar (3x_+^2 - a)}{8\pi x_+} \sqrt{-\Lambda}
\label{temp}
\end{eqnarray}
which has been obtained earlier in \cite{Cai:2009pe,Cai:2009qs,Cai:2009ph} by different methods. It is noted that there exits an extremal limit (i.e. temperature vanishes) at $x_+ = \sqrt{\frac{a}{3}}$, in which $\alpha = 4\Big(\frac{a}{3}\Big)^{3/4}$.

        The next step is to find the entropy of the black hole. As shown in \cite{Cai:2009ph}, the first law of thermodynamics $dM = T_H dS_{BH}$ holds in this case. We shall use this to find the entropy. Using (\ref{mass}) and (\ref{temp}) in the first law of thermodynamics and integrating, we obtain,
\begin{eqnarray}
S_{BH} &=& \frac{\pi p^2\mu^2\Omega_a}{4\hbar}(x_+^2+2a\ln x_+)+S_0
\nonumber
\\
&=&\frac{c^3}{4G}\Big(A - \frac{a\Omega_a}{\Lambda}\ln\frac{A}{A_0}\Big)~.
\label{entropy}
\end{eqnarray} 
Here in the last step, the horizon area $A=\Omega_ar_+^2$ has been used. $A_0$ is the integration constant of dimension of length square. Note that, the entropy is not just proportional to area, as usually happens in Einstein gravity, rather it has an additive term proportional to logarithmic of area. 
\section{Entropy and area spectrum}
    In this section we will derive the spectrum for the entropy as well as the area of the black hole following two methods: tunneling method \cite{Banerjee:2009pf} and QNM method \cite{Hod:1998vk,Kunstatter,Maggiore:2007nq}. Then a comparison of the results obtained in both methods will be done.
\subsection{Tunneling method}
    It has already been mentioned that the pair production occurs inside the horizon. The relevant modes are given by left side of the equality of (\ref{trans1}, \ref{trans2}). 
It has also been shown in the previous section that the left mode is trapped inside the black hole while the right mode can tunnel through the horizon which is observed at asymptotic infinity. 
Therefore, the average value of $\omega$ will be computed as
\begin{eqnarray}
<\omega> = \frac{\displaystyle{\int_0^\infty  \left(\phi^{(R)}_{in}\right)^{*}
\omega  \phi^{(R)}_{in} d\omega}}
{\displaystyle{\int_0^\infty  \left(\phi^{(R)}_{in}\right)^*
\phi^{(R)}_{in} d\omega}} ~.
\label{ref1}
\end{eqnarray}
It should be stressed that the above definition is unique since the pair production occurs inside the black hole and it is the right moving mode that eventually escapes (tunnels) through the horizon.

Since 
the observer is located outside the event horizon,
it is essential to recast the ``in'' expressions into their corresponding ``out'' expressions using the map (\ref{trans1}) and then perform the integrations. 
This yields,
\begin{eqnarray}
<\omega>&=& \frac{\displaystyle{\int_0^\infty e^{-\frac{\pi\omega}{\hbar \kappa}} \left(\phi^{(R)}_{out}\right)^{*}
\omega e^{-\frac{\pi\omega}{\hbar \kappa}} \phi^{(R)}_{out} d\omega}}
{\displaystyle{\int_0^\infty e^{-\frac{\pi\omega}{\hbar \kappa}} \left(\phi^{(R)}_{out}\right)^*
e^{-\frac{\pi\omega}{\hbar \kappa}} \phi^{(R)}_{out} d\omega}}
=\frac{\displaystyle{\int_0^\infty  \omega e^{-\beta\omega}d\omega}}
{\displaystyle{\int_0^\infty  e^{-\beta\omega}d\omega}}
=\beta^{-1}
\label{spec1}
\end{eqnarray}
where $\beta$ is the inverse Hawking temperature
\begin{eqnarray}
\beta=\frac{2\pi}{\hbar \kappa}=\frac{1}{T_H}~.
\end{eqnarray}
In a similar way one can compute the average squared energy of the particle detected by
the asymptotic observer,
$<\omega^{2}>=
\frac{2}{\beta^{2}}$.
Hence
the uncertainty
in the detected energy $\omega$
is given by,
\begin{eqnarray}
\left(\Delta\omega \right)=\sqrt{<\!\!\omega^{2}\!\!>-<\!\!\omega\!\!>^2}\,=\, \beta^{-1} = T_H
\label{spec3}
\end{eqnarray}
which is nothing but the Hawking temperature $T_{H}$.
This uncertainty can be seen as the lack of information in energy of the black hole due to the particle emission.
Now since, as stated earlier, `$\omega$' is the effective energy of the emitted particle as measured by the outside observer and
in the context of information theory, entropy is the lack of information,
then substituting equation (\ref{spec3}) in the first law of black hole mechanics
\begin{eqnarray}
T_{H}(\Delta S_{BH})=\Delta \omega
\label{spec5}
\end{eqnarray}
one obtains
\begin{eqnarray}
\Delta S_{BH}=1 ~.
\label{spec6}
\end{eqnarray}
This shows that the entropy of the black hole is quantized in units of the identity. 
Also, the entropy spectrum is equispaced and
is given by
\begin{eqnarray}
S_n=n
\label{entropyspec1}
\end{eqnarray}
where $n$ is an integer. Since the analysis is semi-classical, the above result is valid only for large $n$. Similar nature of entropy spectrum was obtained in Einstein and Einstein-Gauss-Bonnet theory \cite{Bekenstein:1973ur,Hod:1998vk,Kunstatter,Setare:2003bd,Setare:2004uu,Maggiore:2007nq,Vagenas:2008yi,Daw,Wei,Banerjee:2009pf}.

   A couple of comments are in order here. First, the entropy quantum is universal in the sense that
it is independent of the black hole parameters. This universality was also derived in the
context of the new interpretation of quasi-normal moles of black holes \cite{Maggiore:2007nq, Vagenas:2008yi} for the case of Einstein gravity.
Second, in the Einstein gravity, the same value was also obtained earlier by Hod
by considering the Heisenberg uncertainty principle and Schwinger-type charge emission process \cite{Hod:1999nb}.

\subsection{Quasi-normal mode (QNM) method}     
     In the above analysis, the entropy spectrum is derived by the tunneling mechanism. In the following, we will derive this by the well known method prescribed in \cite{Maggiore:2007nq}. Here the frequency of the quasi-normal modes (QNM) plays an important role. 

  According to this method, a black hole behaves like a damped harmonic oscillator whose frequency is given by $f=(f_R^2 + f_I^2)^{\frac{1}{2}}$, where $f_R$ and $f_I$ are the real and imaginary parts of the frequency of the QNM. In the large $n$ ($n$ is an integer) limit $f_I>> f_R$. Consequently one has to use $f_I$ rather than $f_R$ in the adiabatic quantity \cite{Kunstatter},
\begin{eqnarray}
I_{adiab} = \int\frac{dW}{\Delta f(W)}, \,\,\,\ \Delta f = f_{n+1}-f_n
\label{adiabatic1}
\end{eqnarray}
where $W$ is the energy of the QNM.

      Here we shall calculate the entropy spectrum by using the adiabatic invariant quantity (\ref{adiabatic1}). Since in the large $n$ limit, the imaginary part of the frequency of the QNM is relevant, our next task is to find this. It will be derived by the method prescribed in \cite{Choudhury:2003wd}. For simplicity, we consider a massless scalar field satisfying the wave equation $\nabla^\mu\nabla_\mu\phi = 0$ in the space-time (\ref{solution1}) where the metric coefficients are given by (\ref{solution2}). We look for a solution to this wave equation of the form,
\begin{eqnarray}
\phi=\frac{1}{r}F(r)Y_{lm}(\theta,\phi)e^{\frac{iEt}{\hbar}}
\label{phi}
\end{eqnarray}
with Re($E$)$>0$. Substituting this in the wave equation and performing some simple algebra we obtain the following ``Schr$\ddot{o}$dinger like equation'',
\begin{eqnarray}
\Big[-\frac{d^2}{dr^{*2}}+V(r)\Big]F(r)=\frac{E^2}{\hbar^2}F(r)
\label{Schro}
\end{eqnarray}
where the effective potential $V(r)$ is given by,
\begin{eqnarray}
V(r)=f(r)\Big[\frac{l(l+1)}{r^2}+\frac{f'(r)}{r}\Big]
\label{potential}
\end{eqnarray} 
and the tortoise coordinate $r^*$ is defined as,
\begin{eqnarray}
r^* = \int\frac{dr}{f(r)}.
\label{1.25}
\end{eqnarray}

   In principle (\ref{Schro}) can be solved with a particular set of boundary conditions. But unfortunately, this equation cannot be solved exactly. Therefore to solve this one has to take the help of some approximate method. Here, we shall use the approximation method prescribed in \cite{Choudhury:2003wd}. Note that the effective potential (\ref{potential}) vanishes at the horizon ($r^*\rightarrow -\infty$) and diverges at spatial infinity ($r^*\rightarrow\infty$). Therefore, 
the QNMs are defined to be those for which one has purely ingoing plane wave at the horizon and no wave at spatial infinity, i.e.
\begin{eqnarray}
F(r)|_{QNM} \sim \left\{ \begin{array}{ll}
e^{\frac{iEr^*}{\hbar}} & \textrm{at $r^*\rightarrow -\infty$}\\
0 & \textrm{ at $r^*\rightarrow\infty$}
\end{array} \right.
\label{wave2}
\end{eqnarray}
Now we will solve equation (\ref{Schro}) in the near horizon limit and then impose the above boundary conditions to find the frequency of QNM.

   Expansion of the metric coefficient around the event horizon yields,
\begin{eqnarray}
f(r) &= & f'(r_+)(r-r_+) + {\cal{O}}[(r-r_+)^2]
\nonumber
\\
&=& 2{\kappa}(r-r_+) + {\cal{O}}[(r-r_+)^2]~.
\label{expand}
\end{eqnarray} 
Here in the last step (\ref{sgrav}) has been used. Substituting this in the definition of `$r^*$' (\ref{1.25}) and performing the integration we obtain,
\begin{eqnarray}
r^* &=& \int \frac{dr}{2{\kappa}(r-r_+) + {\cal{O}}[(r-r_+)^2]}
\nonumber
\\
&\simeq& \frac{1}{2\kappa}\ln(r-r_+) + {\cal{O}}[r]
\label{expandto}
\end{eqnarray}
Keeping the first term of `$f(r)$' (\ref{expand}) only and substituting in (\ref{potential}) yields, 
\begin{eqnarray}
V(r)\simeq 2\kappa(r-r_+)\Big[\frac{l(l+1)}{r^2}+\frac{2\kappa}{r}\Big]
\label{nearpotential1}
\end{eqnarray}
Now substituting $\epsilon=r-r_+$ in the above and Taylor expanding around $\epsilon=0$ we obtain the near horizon form of the effective potential:
\begin{eqnarray}
V(\epsilon)\simeq 2\kappa\epsilon\Big[\frac{l(l+1)}{r_+^2}(1-\frac{2\epsilon}{r_+})+\frac{2\kappa}{r_+}(1-\frac{\epsilon}{r_+})\Big]~.
\label{nearpotential2}
\end{eqnarray}
Therefore, keeping only the first term in (\ref{expandto}) and then substituting (\ref{nearpotential2}) in (\ref{Schro}), we obtain the near horizon ``Schrodinger like equation'':
\begin{eqnarray}
-4\kappa^2\epsilon^2\frac{d^2 F}{d\epsilon^2} - 4\kappa^2\epsilon\frac{dF}{d\epsilon}+ 2\kappa\epsilon\Big[\frac{l(l+1)}{r_+^2}(1-\frac{2\epsilon}{r_+})+\frac{2\kappa}{r_+}(1-\frac{\epsilon}{r_+})\Big]F = \frac{E^2}{\hbar^2}F
\label{nearscho}
\end{eqnarray}
Solution of the above equation yields,
\begin{eqnarray}
F \sim  \epsilon^{\frac{iE}{2\hbar\kappa}} U\Big[\frac{1}{4} \Big(2-\frac{i (\frac{\sqrt{\kappa} (2 r_+ \kappa +l+l^2)}{\sqrt{r_+} \sqrt{r_+ \kappa +l+l^2}}-\frac{2 E}{\hbar})}{\kappa}\Big),1+\frac{i E}{\hbar\kappa},\frac{2 i\epsilon \sqrt{r_+ \kappa+l+l^2} }{r{_+}{^{3/2}} \sqrt{\kappa}}\Big]
\label{solve1}
\end{eqnarray}
where `$U[.....]$' is the confluent hypergeometric function. In the limit $\epsilon<<1$, the above solution reduces to the form
\begin{eqnarray}
F &\sim& {\textrm{(constant)}}~\epsilon^{-\frac{iE}{2\hbar\kappa}}\frac{\Gamma(\frac{iE}{\hbar\kappa})}{\Gamma(\frac{1}{2}-\frac{i (\frac{\sqrt{\kappa} (2 r_+ \kappa+l+l^2)}{\sqrt{r_+} \sqrt{r_+ \kappa+l+l^2}}-\frac{2 E}{\hbar})}{4\kappa})}
\nonumber
\\
&+&{\textrm{(constant)}}\epsilon^{\frac{iE}{2\hbar\kappa}}\frac{\Gamma(-\frac{iE}{\hbar\kappa})}{\Gamma(\frac{1}{2}-\frac{i (\frac{\sqrt{\kappa} (2 r_+ \kappa+l+l^2)}{\sqrt{r_+} \sqrt{r_+ \kappa+l+l^2}}+\frac{2 E}{\hbar})}{4\kappa})}~.
\label{solve2}
\end{eqnarray}
In the above expression we did not mention the explicit form of the ``constant'' since in this analysis it is not necessary. The second term of (\ref{solve2}) represents the ingoing wave. Now since there is no outgoing wave in the QNM (\ref{wave2}) at $r^*\rightarrow -\infty$, the first term should vanish. This will happen at poles of the gamma function of the denominator of first term.
The poles of this gamma function will ultimately determine the imaginary part of the frequency of the QNMs. The poles are given by
\begin{eqnarray}
E_n=\Big[1+i(2n+1)\Big]\hbar\kappa
\label{pole}
\end{eqnarray}
for the $l=0$ mode and $n=1,2,3,...$. At these poles the gamma function on the numerator does not vanish. So, these will give the imaginary part of the frequency of the QNM.
Hence, in this case the imaginary part of the frequency of the QNMs is
\begin{eqnarray}
{\textrm{Im}}~ f_n = (2n+1)\kappa = (2n+1)\frac{2\pi T_H}{\hbar},
\label{frequency}
\end{eqnarray}
where $f=\frac{E}{\hbar}$. Of course, one can check this value by the perturbation method as done in \cite{Konoplya:2009ig} for another black hole solution in Ho\v{r}ava-Lifshitz gravity.

   Now the energy of the black hole is given by the ADM mass `$M$' (\ref{mass}) and since from (\ref{frequency}) ${\textrm{Im}}~\Delta f = {\textrm{Im}}(f_{n+1}-f_n) = \frac{4\pi T_H}{\hbar}$, the adiabatic invariant quantity (\ref{adiabatic1}) in this case yields,
\begin{eqnarray}
I_{adiab} = \frac{\hbar}{4\pi}\int\frac{dM}{T_H}~.
\label{adiabatic2}
\end{eqnarray}
Use of first law of thermodynamics, $T_H dS_{BH}=dM$, then leads to,
\begin{eqnarray}
I_{adiab}=\frac{\hbar}{4\pi}\int dS_{BH}=\frac{\hbar}{4\pi}S_{BH}~.
\label{adiabatic3}
\end{eqnarray}
Finally, the Bohr-Sommerfield quantization rule
\begin{eqnarray}
I_{adiab}=n\hbar,
\label{Bohr}
\end{eqnarray}
gives the spacing of the entropy spectrum:
\begin{eqnarray}
S_n= 4\pi n~.
\label{entropyspec2}
\end{eqnarray}

 Although the exact value (\ref{entropyspec1}, \ref{entropyspec2}) of the equi-spacing in the two methods does not coincide, their order of magnitude is same. This discrepancy may be due to the following reason. It has been shown in \cite{Hod:1999nb} that, if one calculates the entropy spectrum following \cite{Bekenstein:1973ur}, by incorporating both the uncertainty relation and the Schwinger mechanism, then the spacing between two adjacent levels is different from the calculation \cite{Bekenstein:1973ur} where the latter effect is not considered. In this connection, one must note that the tunneling mechanism has a similarity with Schwinger mechanism \cite{Paddy,Kim} and also, as we have explained earlier, the uncertainty relation is there. On the contrary, the QNM method incorporates only the uncertainty relation through Bohr-Sommerfield quantization rule (\ref{Bohr}). Hence it not surprising that we obtained a different spacing in the entropy spectrum in both methods.      
  
      Finally, from the expression for the entropy of the black hole given by (\ref{entropy}), we observe that it is not proportional to the area. Therefore, in this case, the area spacing is not equidistant. This is contrary to Einstein gravity but agrees with other examples like Einstein-Gauss-Bonnet gravity.
\section{Conclusions}
    Some aspects of the quantum nature of black holes were studied for the black hole solution recently found in Ho\v{r}ava-Lifshitz gravity. We mainly concentrated on the Hawking effect and black hole spectroscopy. The analysis was done by our reformulated tunneling approach \cite{Banerjee:2008sn,bm2}. The advantage of this reformulated approach is that the emission spectrum was directly obtained instead of just the temperature, as happens in the conventional formulations \cite{Paddy,Wilczek}. Also, as an application, the nature of spectroscopy of the black hole was discussed, following \cite{Banerjee:2009pf}.

     In the semi-classical limit, the analysis showed that the emission spectrum was perfectly black body with a temperature proportional to the surface gravity. This reproduced the familiar form which occurs in known theories, e.g. Einstein and Einstein-Gauss-Bonnet gravities. Using the first law of thermodynamics the entropy was also calculated. The standard Bekenstein-Hawking area law was violated since there was an additional term proportional to logarithmic of area.

    Also, we discussed about the spectrum of entropy/area of the black hole in two distinct ways - the tunneling and QNM approaches. Both revealed that the entropy spectrum was equispaced in the large quantum number limit as usually happens for Einstein gravity and Einstein-Gauss-Bonnet gravity. On the other hand, since the entropy was not proportional to the area, the area spectrum was not equispaced. Consequently, it has a similarity with the Einstein-Gauss-Bonnet theory, rather than the usual Einstein gravity. We hope that the several new results and insights gained from our analysis would help in providing a better understanding of the black holes in Ho\v{r}ava-Lifshitz gravity.

\vskip 5mm
{\bf{Acknowledgement :}}\\
    I thank Prof. Rabin Banerjee for illuminating discussions and a careful reading of the manuscript.

%
%

\end{document}